\documentclass[aps,pre,preprint,groupedaddress,amsmath,amssymb,showpacs,floatfix]{revtex4}
\usepackage{graphicx}
\usepackage{bm}
\providecommand{\mean}[1]{\langle #1 \rangle}
\providecommand{\mean}[1]{{\bm\langle} #1 {\bm\rangle}}
\providecommand{\abs}[1]{\bigl| #1 \bigr|}
\def\vec#1{{\bf{#1}}}

\begin{document}

\title{Brownian motion of a charged particle driven internally by correlated noise}
\author{Francis N.~C.~Paraan}
\email{fcparaan@up.edu.ph}
\affiliation{National Institute of Physics, University of the Philippines, Diliman, Quezon City, Philippines}
\author{Mikhail P.~Solon}
\affiliation{National Institute of Physics, University of the Philippines, Diliman, Quezon City, Philippines}
\author{J.~P.~Esguerra}
\email{pesguerra@nip.upd.edu.ph}
\affiliation{National Institute of Physics, University of the Philippines, Diliman, Quezon City, Philippines}
\date[]{Received 31 August 2007}
\preprint{PREPRINT: TPG041207}

\begin{abstract}
We give an exact solution to the generalized Langevin equation of motion of a charged Brownian particle in a uniform magnetic field that is driven internally by an exponentially-correlated stochastic force. A strong dissipation regime is described in which the ensemble-averaged fluctuations of the velocity exhibit transient oscillations that arise from memory effects. Also, we calculate generalized diffusion coefficients describing the transport of these particles and briefly discuss how they are affected by the magnetic field strength and correlation time. Our asymptotic results are extended to the general case of internal driving by correlated Gaussian stochastic forces with finite autocorrelation times.
\end{abstract}

\pacs{05.40.Jc, 05.40.-a, 05.40.Ca, 05.10.Gg} 

\maketitle

The study of the Brownian motion of a charged particle in uniform and magnetostatic fields is of great importance in the description of the diffusion and transport of plasmas and heavy ions \cite{taylor,behram,behram2,liboff,lemons,czopnik,simoes,jimenez3,jimenez2}. In these references the respective authors modeled the dynamics of the charged particle through the use of a Langevin equation \cite{kampen} of the form 
\begin{equation}\label{langevin}
		{\bf{{\dot{V}}}}(t) + \bm{\gamma}\vec{V}(t) - \omega\vec{V}(t){\bm\times}\frac{\bm{\vec{B}}}{|\vec{B}|}= \bm{\eta}(t),
\end{equation}
where $\bm\gamma$ is generally a constant dissipation rate tensor of rank two, $\omega$ is the cyclotron frequency, \vec{B} is the uniform magnetostatic induction field, and $\bm{\eta}(t)$ is the stochastic driving force per unit mass. Among the main objectives of these works are (i) to obtain expressions for the transport diffusion coefficient from the Langevin equation of motion in the presence of delta-correlated (or white) \cite{taylor, behram, behram2,liboff,czopnik,simoes,jimenez3} and correlated (or colored) noise \cite{lemons,jimenez2}, and (ii) the construction and solution of the associated Fokker-Planck equation \cite{czopnik,simoes,jimenez2,jimenez3}. A drawback of these Langevin models emerges when the charged particle is driven by colored noise so that it generally does not achieve thermal equilibrium with the surrounding bath unless a restrictive condition is imposed on the dissipation rate and the autocorrelation time \cite{luczka}. This situation is undesirable when the particle is driven internally $\bm($such as when the driving fluctuations are mainly due to collisions with particles constituting the fluid bath \cite{kampen}$\bm)$ and physical considerations require us to independently specify the dissipation rate and autocorrelation time. 

In this paper we address this problem by modeling the particle dynamics with a generalized Langevin equation, replacing the dissipation force term of the Langevin equation with one that contains a memory integral and subsequently imposing a generalized fluctuation dissipation relation \cite{kubo,henery}.  In the presence of an exponentially-correlated stochastic driving force, this equation is solved exactly and representative ensemble-averaged trajectories given general initial conditions are presented. The average fluctuations in the position and velocity components are also calculated, allowing us to describe the thermalization and transport of these charged particles in uniform magnetic fields with the appropriate Gaussian noise and continuum approximations. Finally, we use Tauberian theorems to extend our asymptotic results for the case of arbitrary colored Gaussian driving noise with a finite correlation time.

We consider a particle with charge $q$ and mass $m$ under the influence of a uniform magnetic induction field $\vec{B}$ and driven internally by a stochastic force per unit mass $\bm{\eta}(t)$. Since the Lorentz force on the particle is perpendicular to the induction field, the motion of the particle in the direction of this field is governed by a free particle generalized Langevin equation with a well-known solution \cite{mazo}. We therefore restrict the following analysis to the dynamics of the particle projected onto a plane perpendicular to the magnetic induction field $\vec{B}$. Thus, the planar equation of motion expressed in Cartesian coordinates is
\begin{equation}\label{eom}
		{\bf{{\dot{V}}}}(t) + \int_0^t \Gamma(t-t')\vec{V}(t')\,dt' - i\omega\bm{\sigma}_y\vec{V}(t)= \bm{\eta}(t),
\end{equation}
where $\Gamma(t)$ is the memory dissipation kernel, $\omega$ is the cyclotron frequency $qB/m$, and $\bm{\sigma}_y$ is the Pauli matrix $\bigl(\begin{smallmatrix}0&-i\\i&0\end{smallmatrix}\bigr)$.

We assume that the components of the stochastic force vector $\eta_i(t)$ have Gaussian distributions, vanishing average values $\mean{\eta_i(t)} = 0$, vanishing cross-correlation functions, and identical exponential autocorrelation functions
\begin{equation}\label{correl}
	\mean{\eta_i(t)\eta_j(t')}  = \delta_{i,j}\frac{\gamma k_BT}{m\tau}e^{-|t-t'|/\tau} = C{\bm(}|t-t'|{\bm)}.
\end{equation}
Here, $\gamma$ is the characteristic dissipation decay rate, $\tau$ is the average correlation time, $k_B$ is Boltzmann's constant, and $T$ is the absolute temperature of the bath surrounding the charged particle.  These properties are based on the assumptions that (i) the bath is rotationally invariant about the field axis, (ii) particle collisions are accurately modeled as an Ornstein-Uhlenbeck process \cite{luczka}, and (iii) inasmuch as the observation time is much shorter than the correlation time, a Fokker-Planck equation with a generalized diffusion coefficient proportional to the fluctuations in position can be effectively constructed.

To completely specify our problem, we finally apply the condition of internal driving by imposing a generalized fluctuation-dissipation relation \cite{kubo} on the dissipation memory kernel and find that 
\begin{equation}
\Gamma(t) = \frac{\gamma}{\tau}e^{-t/\tau}, \quad t>0.	
\end{equation}

We begin our solution by obtaining the Laplace transform of the velocity
\begin{equation}\label{magxs}
 \tilde{\vec{V}}(s) = \frac{1}{s+\tilde\Gamma}\biggl[ \vec{I} - \frac{i\omega\bm{\sigma}_y}{s+\tilde\Gamma}\biggr]^{-1}[\vec{V}_0 + \tilde{\bm{\eta}}(s)],
\end{equation}
where $\vec{I}$ is an identity matrix, $\vec{V}_0$ is the initial velocity vector, and $\tilde{\Gamma}(s)$ and $\tilde{\bm{\eta}}(s)$ are the transforms of the dissipation memory kernel $\Gamma(t)$ and the driving stochastic term $\bm{\eta}(t)$, respectively.  The inverse matrix appearing in this transform may be formally expressed as a geometric series giving us
\begin{equation}
 \tilde{\vec{V}} =\frac{1}{s+\tilde\Gamma}\sum_{n=0}^\infty \biggl(\frac{i\omega\bm{\sigma}_y}{s+\tilde\Gamma}\biggr)^n[\vec{V}_0 + \tilde{\bm{\eta}}].
\end{equation}
We split the resulting series into partial sums involving even and odd powers of ${\bm\sigma}_y$ and use the fact that ${\bm\sigma}_y^2 = \vec{I}$ to obtain
\begin{equation}\label{exact}
  \tilde{\vec{V}} = \frac{(1+\tau s)\bigl[(\tau s^2+s+\gamma)\vec{I}+i\omega(1+\tau s){\bm\sigma}_y\bigr]}{(\tau s^2+s+\gamma)^2+\omega^2(1+\tau s)^2}\,[\vec{V}_0 + \tilde{\bm{\eta}}].
\end{equation}

\begin{figure}[tb]
\begin{center}
   \includegraphics[width=1\linewidth,clip]{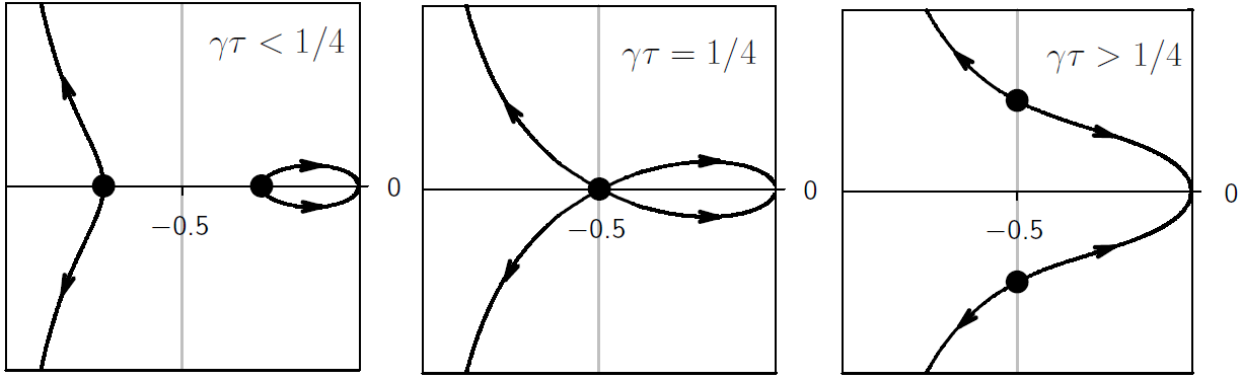} 
\end{center}
\caption{Poles of the transfer function described in the text for non-negative $\omega\tau$. Black dots represent the locations of these poles when $\omega\tau = 0$ and arrowheads denote the direction of increasing $\omega\tau$.\label{figpoles}}
\end{figure}

We identify the four poles of the transfer function multiplying $[\vec{V}_0 + \tilde{\bm{\eta}}]$ as 
\begin{equation}
	s_\pm = \frac{-(1+i\omega\tau)\pm\sqrt{1-4\gamma\tau-\omega^2\tau^2 + i2\omega\tau}}{2\tau},
\end{equation}
and their complex conjugates $s_\pm^*$. The locations of these poles in the complex plane are shown in Fig.~\ref{figpoles}. A brief examination of this pole structure suggests that there exists a regime $\gamma\tau\lesssim 1/4$ where transient oscillations in the velocity components are negligible on the time scale of $\tau$ for small cyclotron frequencies. Indeed, requiring the dominant relaxation time to be much smaller than the characteristic period of these oscillations under such a condition leads to the inequality $\omega \ll \gamma$. On the other hand, when the product $\gamma\tau$ is much larger than $1/4$ these transient oscillations remain observable even as the cyclotron frequency vanishes, implying that they result from the memory of the system.  This dramatic behavior will be demonstrated below as we construct the exact solutions to our generalized Langevin equation.

With the poles of the transfer function identified, we can easily calculate the Laplace inverse of $\tilde{\vec{V}}$ by an integration over the appropriate Bromwich contour. The desired solution is
\begin{equation}
	\vec{V}(t) = \bigl[g_1(t)\vec{I} + ig_2(t){\bm\sigma}_y\bigr]{\bm\ast}\bigl[\vec{V}_0\delta(t) + {\bm{\eta}}(t)\bigr],
\end{equation}
where ${\bm\ast}$ is the Laplace convolution operator \footnote{The action of the Laplace convolution operator is given by $f(t){\bm\ast} g(t) = \int_0^t f(t')g(t-t')\,dt'$.}, $\delta(t)$ is the usual Dirac delta function, and $g_1(t)$ and $g_2(t)$ are the real and imaginary parts, respectively, of the complex-valued function
\begin{equation}
	g(t) = e^{-(1-i\omega\tau)t/2\tau}\biggl[\cosh \frac{\beta t}{2\tau} +\frac{1+i\omega\tau}{\beta}\,\sinh \frac{\beta t}{2\tau} \biggr].
\end{equation}
In the preceding expression, we have made the substitution $\beta \equiv \sqrt{1-4\gamma\tau-\omega^2\tau^2 + i2\omega\tau}$.	
Since the stochastic force has a vanishing mean, we find that the ensemble-averaged velocity of the particle is
\begin{equation}
	\mean{\vec{V}(t)} = \bigl[g_1(t)\vec{I} + g_2(t)i{\bm\sigma}_y\bigr]\vec{V}_0.
\end{equation}
The average velocity asymptotically decays to zero as expected because of the presence of friction and the fact that the magnetic field does no work on the particle. The relaxation to equilibrium is generally damped biexponentially with characteristic times $2\tau/{\bm(}1\pm \mathfrak{Re}[\beta]{\bm)}$. 

By setting the origin of the Cartesian axes at the initial position of the particle and integrating $\mean{\vec{V}(t)}$ we are able to calculate the average position of the Brownian particle as
\begin{equation}\label{traj}
	\mean{\vec{X}(t)} = \bigl[G_1(t)\vec{I} + G_2(t)i{\bm\sigma}_y\bigr]\vec{V}_0,
\end{equation}
where $G_1(t)$ and $G_2(t)$ are the real and imaginary parts, respectively, of the function
\begin{equation}
G(t) = \frac{e^{-(1-i\omega\tau)t/2\tau}}{\gamma-i\omega}\biggl[ e^{(1-i\omega\tau)t/2\tau}-\cosh\frac{\beta t}{2\tau}  -\frac{1-2\gamma\tau+i\omega\tau}{\beta}\sinh\frac{\beta t}{2\tau}\biggr].
\end{equation}

\begin{figure}[tb]
  \centering
   \includegraphics[width=0.49\linewidth,clip]{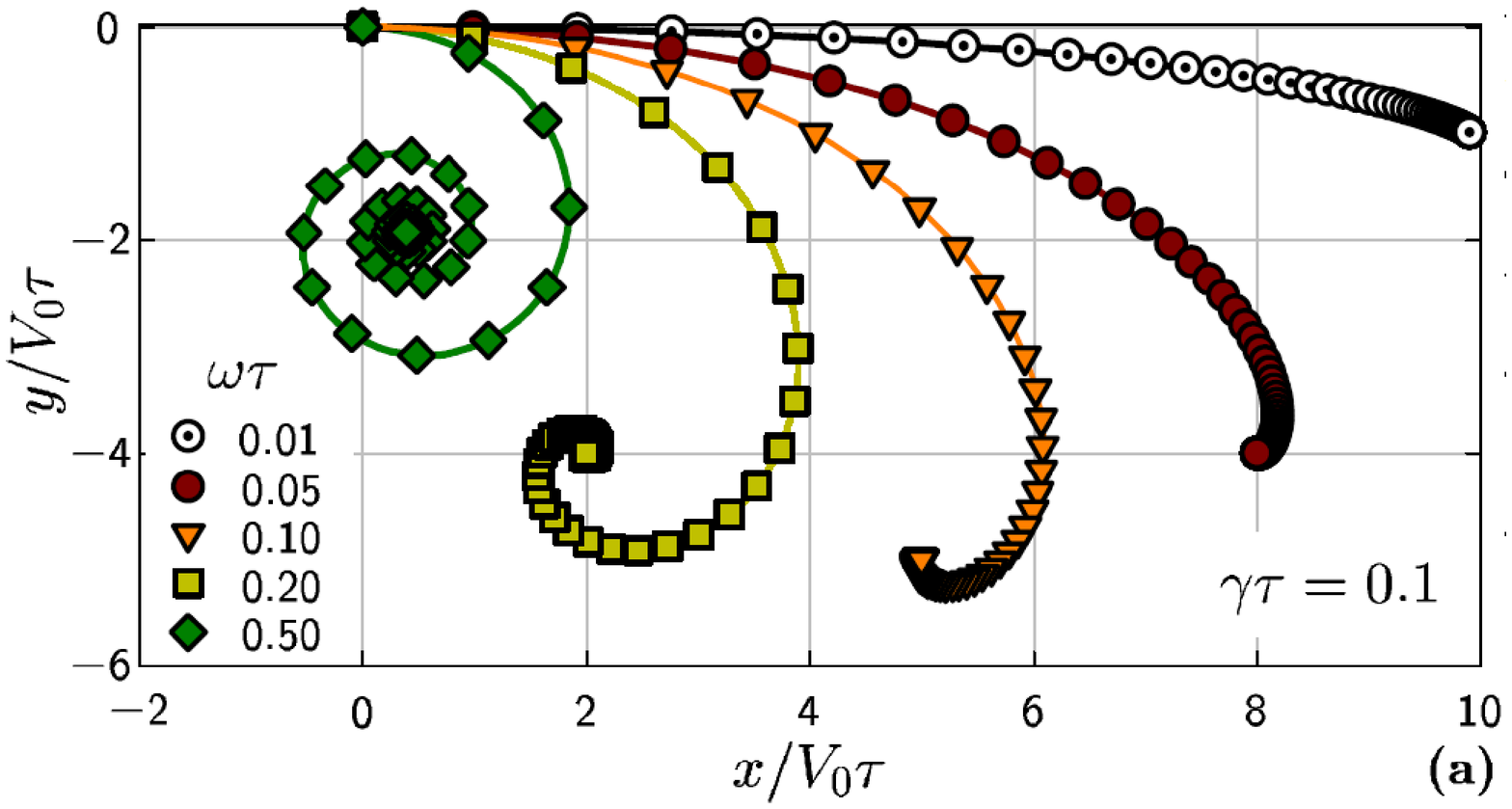} 
   \includegraphics[width=0.49\linewidth,clip]{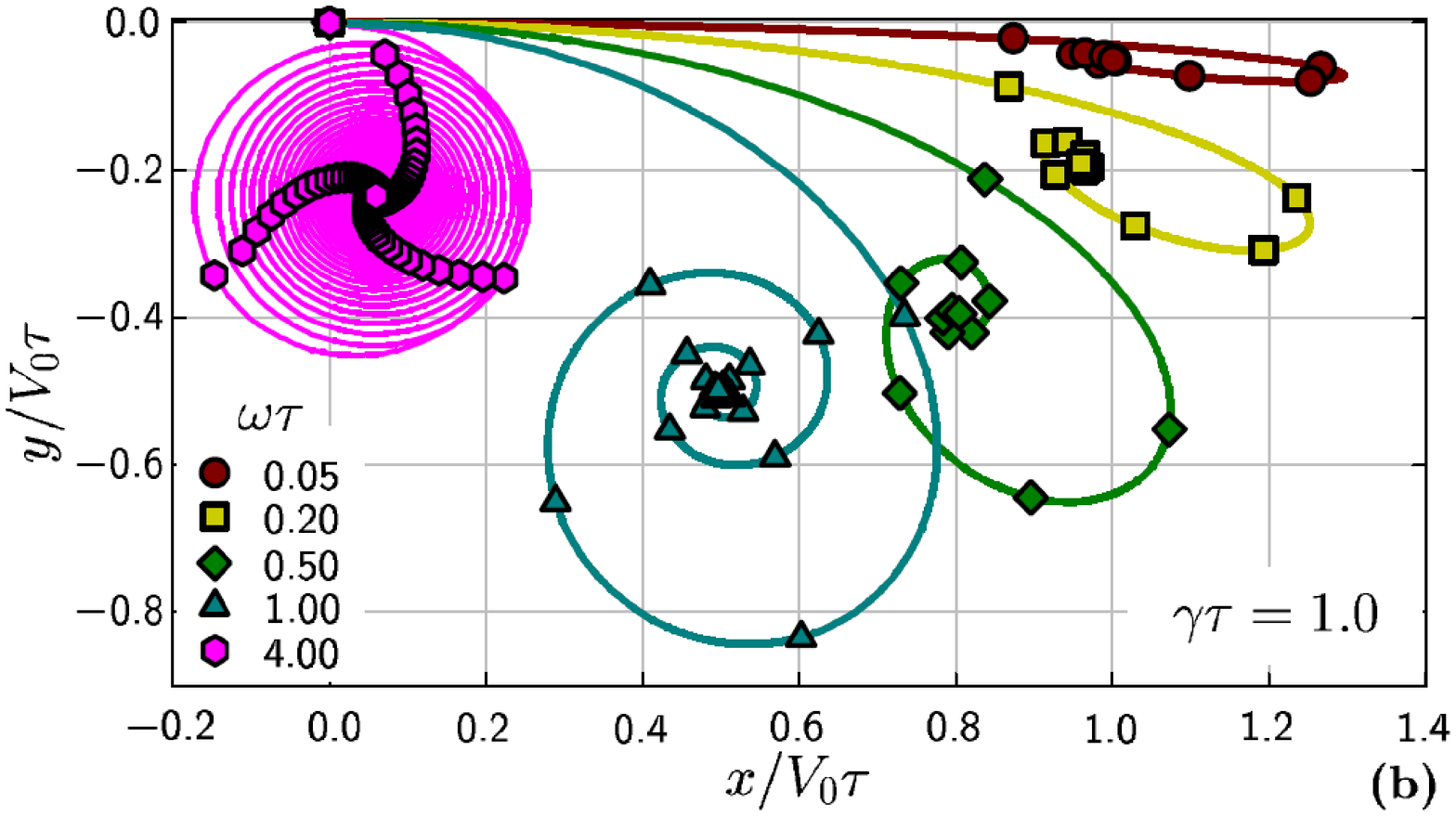} 
\caption{(Color online) Ensemble-averaged trajectories of internally driven charged particles in a uniform magnetic field directed in the $z$-direction. The product $\gamma\tau$ is equal to $0.1$ (a) and $1.0$ (b). Adjacent symbols on each curve are separated in time by an interval $\tau$.\label{gt1}}
\end{figure}

Representative ensemble-averaged particle trajectories that were calculated from Eq.~\ref{traj} with the initial velocity $\vec{V}_0$ in the $x$-direction are depicted in Fig.~\ref{gt1}. Consistent with the preceding analysis of the pole locations of the transfer function, we see that the average velocity components depict marked transient oscillations even in vanishing magnetic fields when the dissipation rate is high ($\omega\tau =0.05$ in Fig.~\ref{gt1}b). The long-time equilibrium position of the charged particle relative to its original position is
\begin{equation}
	\mean{\vec{X}^\infty} = \begin{pmatrix}\gamma & \omega\\-\omega & \gamma\end{pmatrix}\frac{\vec{V}_0}{\gamma^2 + \omega^2}.
\end{equation}
As this expression is independent of the autocorrelation time, this result may be used to experimentally deduce the average dissipation rate $\gamma$ of appropriately prepared charged particle systems.

The fluctuations in the position and velocity components may be expressed in terms of the functions $g(t)$ and $G(t)$ by noting the independence and equivalence of the noise components (Eq.\ \ref{correl}) and extending the approach outlined in Refs.\ \cite{porra,mazo} yielding
\begin{align}
	\mean{(\Delta v_i)^2}(t) &= \frac{k_BT}{m}\bigl( 1-\abs{g(t)}^2\bigr), \label{vsquared}\\
	\mean{(\Delta x_i)^2}(t) &= \frac{k_BT}{m}\biggl(2\negthickspace\int_0^t\negthickspace\mathfrak{Re}\bigl[G(t')\bigr]\,dt' - \abs{G(t)}^2\biggr). \label{posfluct}
\end{align}
Evaluating the remaining integral in the last equation leads to a lengthy expression, but the end result is a function consisting of a term linear in time plus transient terms, $\mean{(\Delta x_i)^2} = 2k_BT\gamma t/m(\omega^2 + \gamma^2)\,+\,\text{transient terms}$. This asymptotic linear dependence of the fluctuation in position on time tells us that the associated diffusion process is normal, as we discuss below.

\begin{figure}[tb]
\centering
   \includegraphics[width=0.557\linewidth,clip]{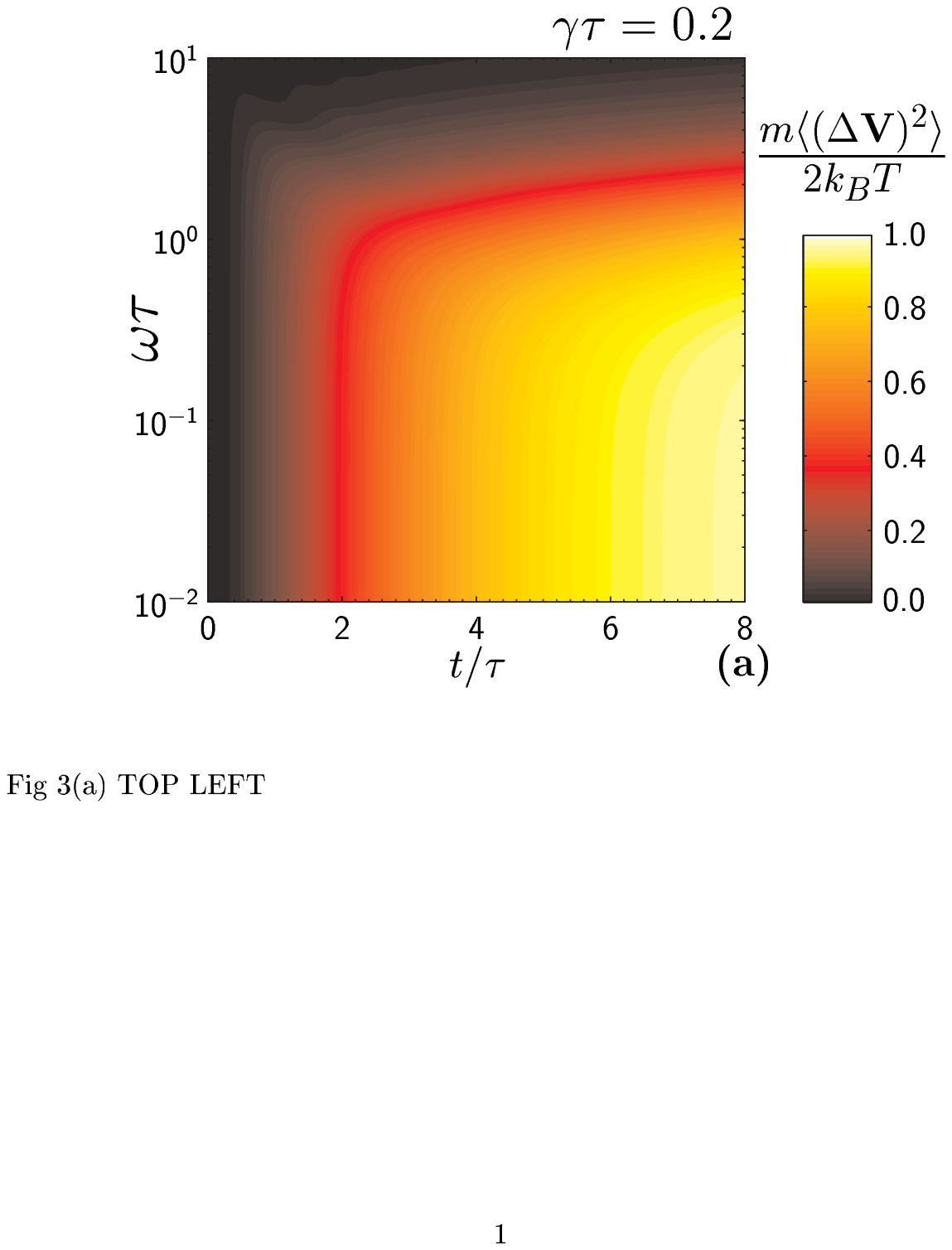}
   \includegraphics[width=0.423\linewidth,clip]{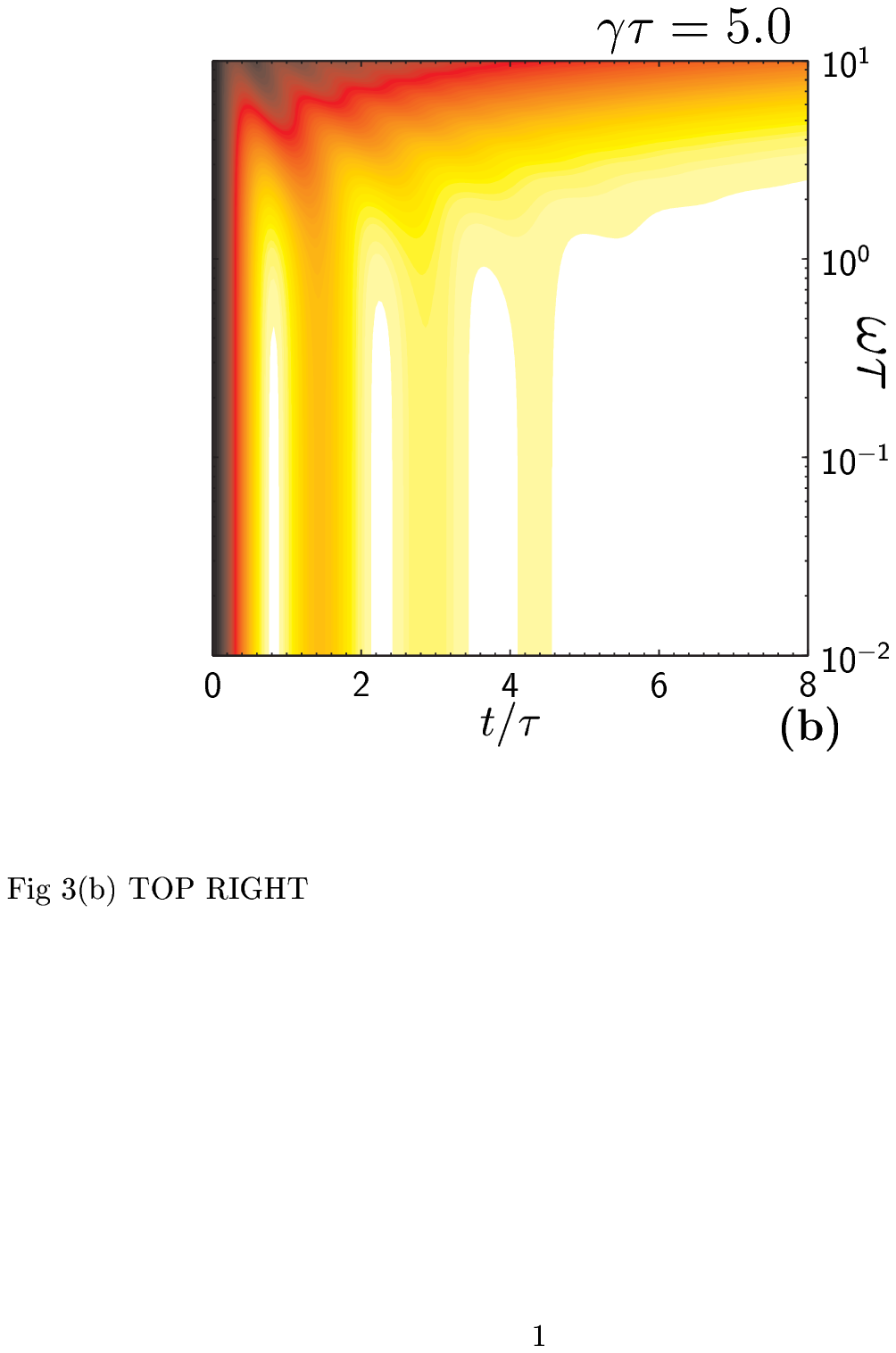}
   \includegraphics[width=\linewidth,clip]{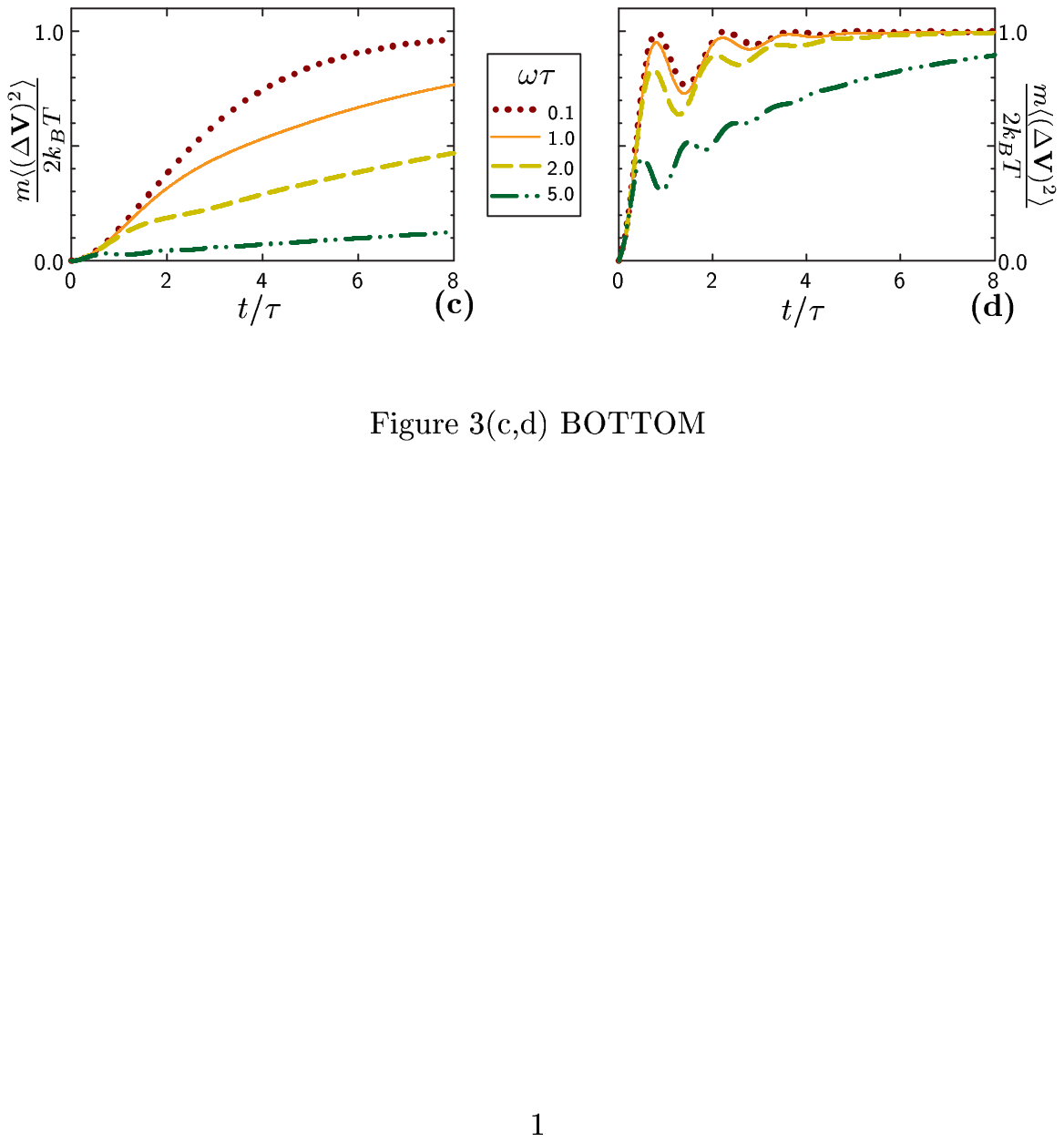}
\caption{(Color online) Time evolution of the velocity fluctuation ${\bm\langle}(\Delta\vec{V})^2{\bm\rangle}$ of the charged particle in units of its asymptotic value $2k_BT/m$. The product $\gamma\tau$ is equal to $0.2$ (a,~c) and $5.0$ (b,~d). \label{vsqr}}
\end{figure}

The fluctuation in velocity may be calculated from the mean-square velocity components (Eq.~\ref{vsquared}) and is graphed in Fig.~\ref{vsqr}.  For the special case of a particle initially prepared with zero velocity, this quantity also corresponds to the average kinetic energy of the charged particle due to its projected motion on a plane perpendicular to the magnetic field \footnote{This average kinetic energy may be calculated from $\tfrac{1}{2}m\sum_i\mean{(\Delta v_i)^2}+\mean{v_i}^2$.}. Since we have imposed the condition of internal driving and the mean velocity vanishes at long times, the quantity $\tfrac{1}{2}m{\bm\langle}(\Delta\vec{V})^2{\bm\rangle}$ eventually relaxes to the expected equipartition energy $k_BT$. In general, the typical time scale $\theta$ of this relaxation is 
\begin{equation}
\theta \sim \max\biggl[\frac{\tau}{1 \pm \mathfrak{Re}[\beta]}\biggr],	
\end{equation}
which scales as $\theta\sim\omega^2\tau^2/\gamma$ for large cyclotron frequencies and small $\gamma$. Also, as we have discussed earlier, the approach of the velocity fluctuation to its equilibrium value for $\gamma\tau<1/4$ and small values of $\omega\tau$ is essentially monotonic, while transient oscillations may be observed for small values of $\omega\tau$ in the presence of strong friction (Figs.~\ref{vsqr}b,~\ref{vsqr}d).

\begin{figure}[tb]
\centering
   \includegraphics[width=0.467\linewidth,clip]{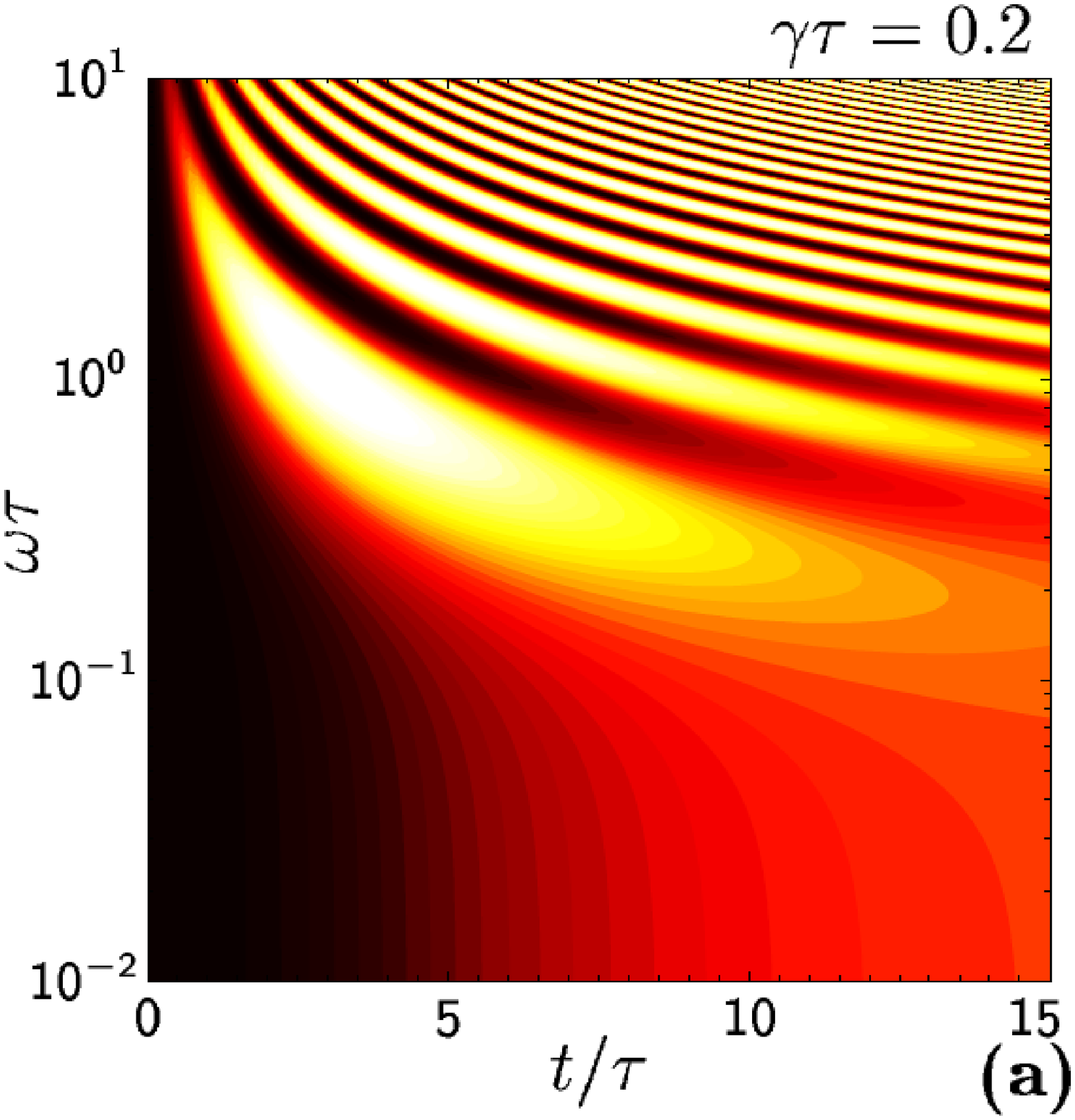}
   \includegraphics[width=0.516\linewidth,clip]{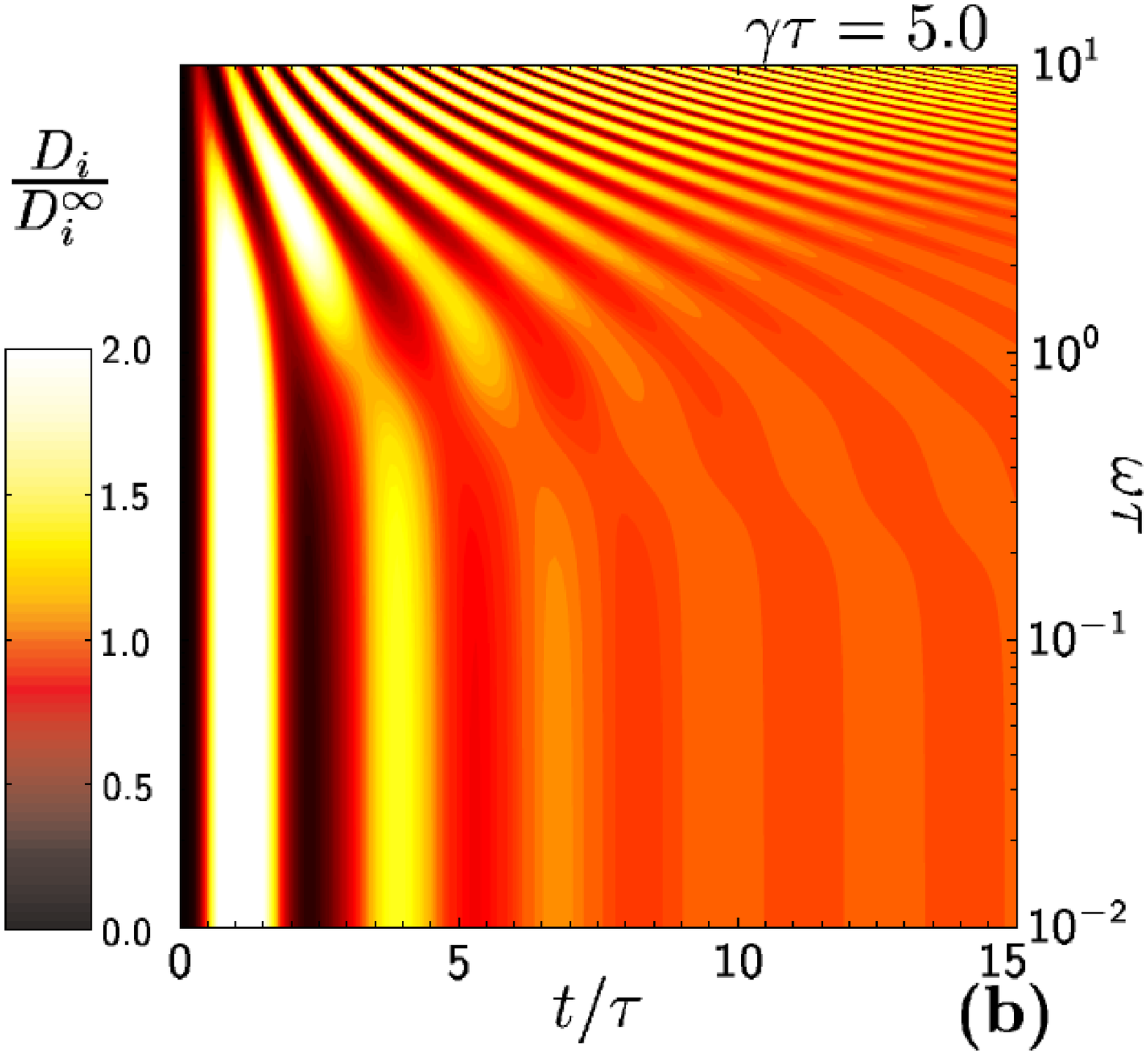}
   \includegraphics[width=\linewidth,clip]{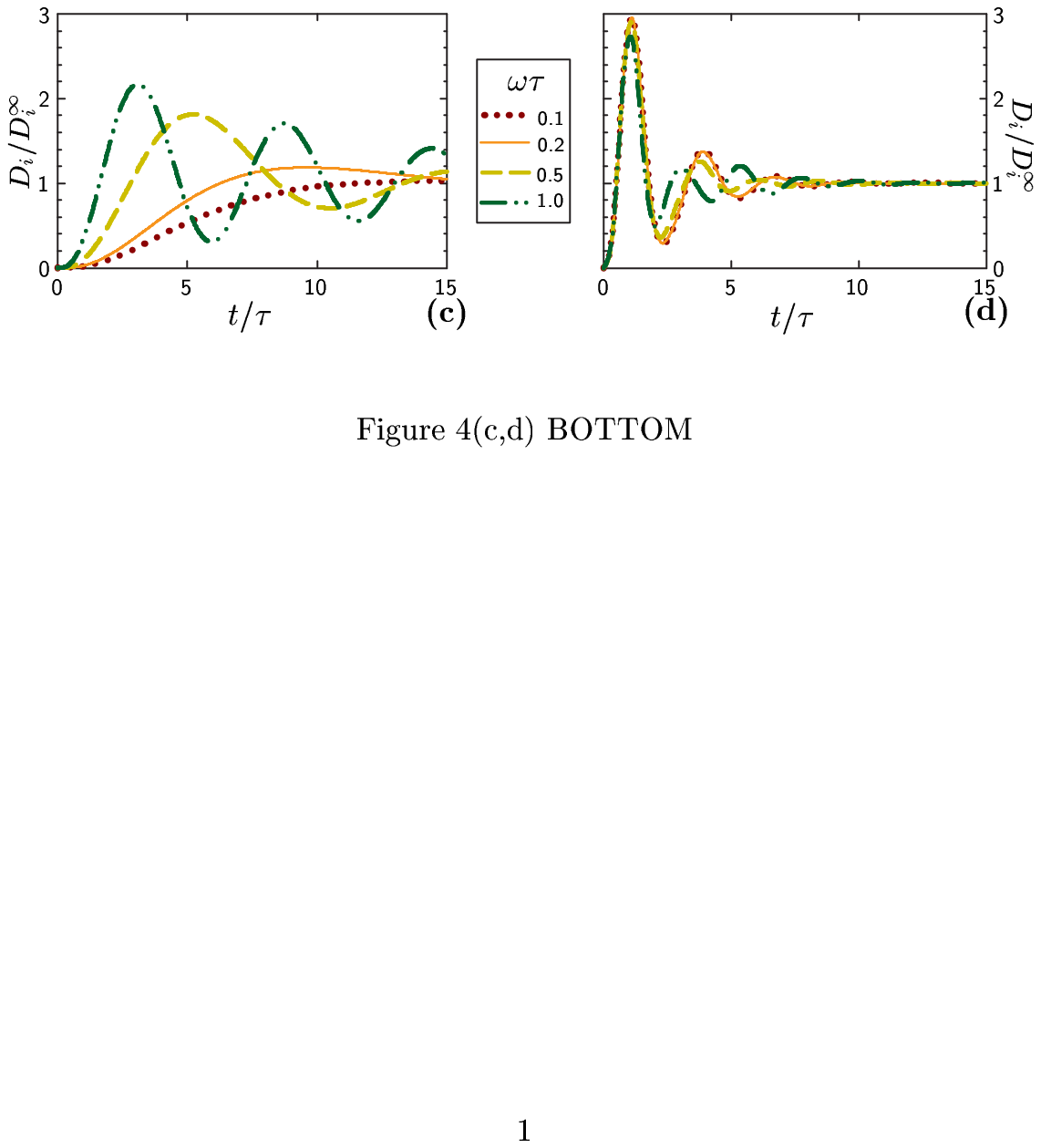}
\caption{(Color online) Time evolution of the one-dimensional diffusion coefficient $D_i$ in units of its asymptotic value $D_i^\infty$. The product $\gamma\tau$ is equal to $0.2$ (a,~c) and $5.0$ (b,~d). \label{dc1}}
\end{figure}

The effective one-dimensional transport diffusion coefficient $D_i(t)$ can be obtained from the fluctuations in the components of the position vector (Eq.~\ref{posfluct}) through the definition $D_i(t) \equiv \tfrac{d}{dt}\mean{(\Delta x_i)^2}$. Calculating this derivative leads to
\begin{equation}
	D_i(t) = \frac{2k_BT}{m}\mathfrak{Re}\bigl\{G(t){[}1-\bar{g}(t){]} \bigr\},
\end{equation}
where the overbar refers to the complex conjugate. In the limit of long times we find that these diffusion coefficients approach the constant value
\begin{equation}
	D_i^\infty = \frac{2k_BT}{m}\frac{\gamma}{\omega^2 + \gamma^2},
\end{equation}
which is independent of the correlation time $\tau$ and correctly gives the zero-field value when $\omega\to 0$.  This general result is reproduced by the analogous Langevin model \cite{lemons,jimenez2} only when the special condition $\gamma\tau=1$ is imposed. Graphs displaying the transient characteristics of the diffusion coefficient are shown in Fig.~\ref{dc1}. In general, an increasing dissipation rate decreases the typical equilibration time as expected (Fig.~\ref{dc1}c,~\ref{dc1}d), although for large $\omega\tau$ the coefficient oscillates strongly before approaching its asymptotic value.  The transient memory-induced oscillations of the diffusion coefficient in the strong friction regime are clearly seen when $\gamma\tau = 5.0$ (Fig.~\ref{dc1}b).

We now turn our attention to generalizing our results to the case of a driving force with a finite characteristic correlation time but with an otherwise arbitrary autocorrelation function. After imposing a fluctuation-dissipation condition as we have done above, Tauberian theorems allow us to extract the long-time behavior of such an autocorrelation function and the associated dissipation memory kernel by a leading-order truncation of the corresponding Laplace transforms \cite{feller}. This approximation, which is valid for small $s$ and large times, reproduces the same Laplace transform solution (Eq.\ \ref{exact}) upon a trivial rescaling of time.  Thus, all the equilibrium values calculated here are generally applicable so long as the stochastic driving process is characterized by a finite autocorrelation time.

In this paper we have analyzed the Brownian motion of a charged particle in the presence of a static and uniform magnetic field and driven internally by an exponentially-correlated stochastic force using a generalized Langevin model. The dynamics of the particle was described exactly when the driving noise is exponentially correlated in time and the velocity fluctuation and diffusion coefficient were calculated for all times. Of particular interest was a strong dissipation regime $\gamma\tau\gg 1/4$ that displayed damped oscillations in these quantities even in the limit of vanishing cyclotron frequencies, which suggests that this transient behavior is memory-induced. Also, equilibrium values of these dynamical quantities were presented and demonstrated to be independent of the characteristic correlation time of the driving noise as required. With the use of Tauberian theorems we have shown that these equilibrium results do not depend on the specific form of the correlation function of the driving force so long as it possesses a finite autocorrelation time.

J.~P.~E. acknowledges the support of a National Institute of Physics Faculty Grant. Part of this work was carried out by F.~N.~C.~P. at The Abdus Salam International Centre for Theoretical Physics, Trieste, Italy.

\end{document}